\documentclass[aps,prb,twocolumn,superscriptaddress,floatfix,a4]{revtex4}
\usepackage{epsfig,amsmath,amssymb,color}
\begin{document}

\title{Non-equilibrium electronic transport in a one-dimensional Mott insulator}

\author{F.~Heidrich-Meisner}
\affiliation {Physics Department, Arnold Sommerfeld Center for Theoretical
Physics, and Center for NanoScience, Ludwig-Maximilians-Universit\"at
M\"unchen, D-80333 M\"unchen, Germany}
\email[Corresponding author: ]{heidrich-meisner@lmu.de}

\author{I.~Gonz\'alez}         
\affiliation {Centro de Supercomputaci\'on de
Galicia, Avenida de Vigo s/n, E-15705 Santiago de Compostela, Spain}

\author{K.~A.~Al-Hassanieh}
\affiliation {Theoretical Division, Los Alamos National Laboratory, Los Alamos, New Mexico 87545, USA}

\author{A.~E.~Feiguin}        
\affiliation {Department of Physics and Astronomy, University of Wyoming, Laramie, Wyoming 82071, USA}

\author{M.~J.~Rozenberg} 
\affiliation{Laboratoire de Physique des Solides, CNRS-UMR8502, Universite de
Paris-Sud, Orsay 91405, France}
\affiliation {Departamento de F\'{\i}sica, FCEN, Universidad de Buenos Aires, Ciudad Universitaria 
Pabell\'on I, Buenos Aires 1428, Argentina} 

\author{E.~Dagotto}     
\affiliation {Materials Science and Technology Division, Oak Ridge National Laboratory, Oak Ridge, 
TN 37831 and Department of Physics,  University of Tennessee, Knoxville, TN 37996, USA}

\begin{abstract}

We calculate the non-equilibrium electronic transport properties of a
one-dimensional interacting chain at half filling, coupled to
non-interacting leads. The interacting chain is initially in a Mott
insulator state that is driven out of equilibrium by applying a strong
bias voltage between the leads. For bias voltages above a certain
threshold we observe the breakdown of the Mott insulator state and the
establishment of a steady-state electronic current through the system. Based on extensive
time-dependent density matrix renormalization group simulations, we
show that this steady-state current always has the same functional
dependence on voltage, independent of the microscopic details of the model
and we relate the value of the threshold to the Lieb-Wu gap. We frame our
results in terms of the Landau-Zener dielectric breakdown picture. Finally, we
also discuss the real-time evolution of the current, and characterize the
current-carrying state resulting from the
breakdown of the Mott insulator by computing the double occupancy, the
spin structure factor, and the entanglement entropy. 
 
\end{abstract}


\maketitle

%%%%%%%%%%%%%%%% Introduction%%%%%%%%%%%%%%%%%%%%%%%%%%%%%%%%%%%%%%%%%%

\section{Introduction}

The theoretical understanding of the non-equilibrium transport properties
of strongly interacting systems in low dimensions has become a very active
field of research, mainly due to  the  experimental activity in the fields
of nanoscale materials~\cite{wiel00,grobis08,scott09} and cold atomic
gases,\cite{bloch08} as well as due to  advances in theoretical methods
designed to deal with both the non-equilibrium situation {\it and}
electronic correlations (see
Refs.~\onlinecite{eckel10,andergassen10,eckstein10} for an overview and
references therein).  When considering non-equilibrium electronic
transport, we have in mind a nanostructure that is subject to a large
external voltage such that linear response theory does not apply anymore.
The main theoretical question that one would like to address is the
dependence of the electrical current on the applied voltage, i.e., 
the current-voltage characteristics, understanding
not only the steady-state current reached on large time-scales, but also
the transient regime appearing on shorter time-scales. Another important
question is the characterization of the current-carrying state,
contrasting its properties against equilibrium states in the absence of a
voltage.  From the experimental point of view, knowledge of the full
dependence of the electronic current on the bias voltage through an
interacting nanostructure is a question of utmost importance, as this
measurement is a standard technique to map out electronic energy levels
and to observe many-body effects in nanostructures (see, e.g.,
Ref.~\onlinecite{vanderwiel02} for experimental work and
Ref.~\onlinecite{alhassanieh09} on theoretical work).

A paramount issue when studying transport in strongly interacting systems
is the behavior of the insulating states characteristic of these systems,
the most relevant of which is the Mott insulator (MI) state. Considerable
theoretical efforts have so far been devoted to the study of
non-equilibrium transport in  nanostructures such as quantum dots (see,
e.g.,
Refs.~\onlinecite{anders08,boulat08,feiguin08c,weiss08,hm09b,sela09,werner10,chao10,jakobs10,karrasch10,muehlbacher10,schiro10}).
Using state-of-the-art-numerical approaches, substantial progress has been made in calculating the current-voltage
characteristics and non-equilibrium properties of some basic models, such as the interacting resonant
level model~\cite{boulat08} or the single-impurity Anderson
model\cite{anders08,hm09b,werner10,muehlbacher10,schiro10} as well as in understanding their transient
behavior.\cite{schmidt08,hm09b,branschaedel10} Whereas quantum dots
with an odd number of electrons exhibit perfect conductance in the low
bias regime due to the Kondo effect,\cite{hewson} an extended 
region with repulsive interactions, an even number of electrons, and at half filling is an
insulator. The crossover from single quantum dots to this Mott insulating
state has been studied in Refs.~\onlinecite{oguri97,oguri99,oguri01} on
the level of linear response theory, showing that the ground state
alternates between a conducting state for an odd number of sites and an
insulating state for an even number of sites.  Of course, in the limit of
large systems, the difference between $N$ and $N+1$ electrons becomes
irrelevant and the perfect conductance in a system with an odd number of
electrons can only be observed at, with respect to experiments,
unrealistically low energy scales.

\begin{figure}
\epsfxsize=7cm \centerline{\epsfbox{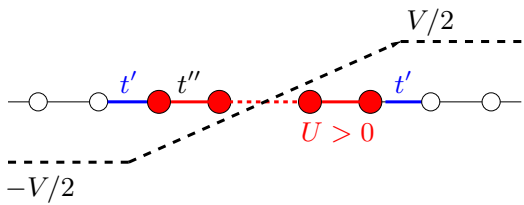}}
\caption{\label{fig:model} 
(Color online) Sketch of the one-dimensional nanostructure
described in the text, with an extended interacting region connected to
non-interacting leads. Open (solid) symbols represent non-interacting
(interacting) sites. The dashed line shows the voltage profile used to
drive the system out of equilibrium. The voltage is homogeneous in the leads,
and interpolates linearly between those values in the
interacting region.  
}
\end{figure}

In this paper, we shall thus turn our attention to non-equilibrium
electronic transport through  an extended interacting region, described 
by the one-dimensional Hubbard model with repulsive onsite interactions.
Specifically we consider a one-dimensional (1D) system consisting of an
interacting region of length $L_{\rm int}$ connected to two non-interacting
leads (see Fig.~\ref{fig:model}). The interacting region is initially in
the MI state and we focus on the strongly interacting regime with interaction 
strength on the order or larger than the bandwidth.  
The sudden application of a large external voltage drives
the system out of equilibrium and causes a time-dependent electrical
current to flow through the interacting region, destroying the MI state.
Our goals are first, to calculate the nonlinear current-voltage (I-V)
characteristics; second, to  contribute to the characterization of the
current-carrying state; and third, to study the time-dependence of the
entanglement entropy in this set-up. We employ the adaptive time-dependent
density-matrix renormalization group (tDMRG) method,\cite{white04,daley04}
which has been successfully used to compute the non-equilibrium dynamics of
single quantum dots.\cite{boulat08,hm09b,dasilva08,kirino08,branschaedel10}

The problem of destroying a MI by subjecting it to a voltage or an
electric field is currently attracting significant attention, both for the
1D \cite{oka03,oka05a,oka05b,dutta07,dutta08,oka10,kirino10} and the 3D
cases,\cite{eckstein10a} as well as in
heterostructures.\cite{okamoto07,okamoto08} In the case of an  extended
interacting region, the voltage can be applied  with different spatial
profiles in theoretical simulations, giving rise to different physical
mechanisms for destroying the MI state.  In the set-up sketched in
Fig.~\ref{fig:model}, we linearly interpolate between the voltages set in
the leads across the interacting region.  As we shall argue, this gives
rise to a many-body Landau-Zener mechanism through which the MI breaks
down. This picture has been advocated for in a series of studies by Oka
and collaborators,\cite{oka03,oka05a,oka05b,oka10} who  considered both a
ring geometry pierced by a time-dependent flux\cite{oka03} and a MI
subject to a linear potential without including a coupling to
leads.\cite{oka05b} Both approaches  model the application of an electric
field. One of their main results is that the breakdown of a
MI is governed by the same  physical laws as the one of a band insulator,
with the difference that the band gap needs to be replaced by the charge
gap of the strongly-interacting MI.\cite{oka03,oka05b} 

Our setup is chosen to  closely catch features of an actual transport
experiment by including the leads. We shall provide a qualitative
comparison of our results with other cases recently addressed in the
literature.\cite{oka05b,oka10,kirino10,eckstein10a} Transport through
extended interacting regions that are not necessarily in a MI state has
been studied as well in
Refs.~\onlinecite{prosen09,michel08,benenti09,benenti09a,mierzejewski10},
emphasizing as recurring themes the appearance of nonlinear
current-voltage characteristics and negative differential conductances.

Our main result is the accurate numerical calculation of steady-state
currents for the geometry of Fig.~\ref{fig:model}. We find that the
current-voltage characteristics can be described by an expression of the
form \begin{equation} \label{eq:iv} J(V) = aVe^{-V_c/{V}}, \end{equation}
in agreement with Refs.~\onlinecite{oka03,oka05a,oka05b,oka10}, implying
that at sufficiently large voltages, the system is driven to a conducting
state with $J\propto V$. We show that $V_c\propto \Delta_c^2$, where
$\Delta_c$ is the Mott gap.  In addition, we analyze several quantities in
the current-carrying state, with a particular focus on the double
occupancy and spin correlations. While the current-carrying state still
has a tendency towards antiferromagnetic correlations, this instability is
strongly suppressed compared to the MI state. However, neither the
spin-structure factor nor the double occupancy, which is a measure of the
interaction energy stored in the interacting region, saturate in the time
window that we can access numerically. This suggests that the interacting
region still undergoes a reorganization of internal energy while the
particle flow in and out of the interacting region is already constant.
The crossover from the insulating regime to the conducting regime is also
reflected in the time-dependence of the entanglement entropy. We further
show that this quantity behaves similarly to the case of global quenches:
in our set-up, which is relevant for transport, the entanglement entropy
increases linearly in time in the conducting regime.  Here, the increase
of entanglement is due to real particles moving around, different from the
situation encountered in quantum quenches with homogeneous particle
densities, in which propagating collective excitations induce
entanglement.\cite{calabrese05,calabrese07}

One-dimensional Mott insulators can be realized experimentally in several
classes of materials.  A promising class of materials that have been
suggested to realize 1D MI are carbon
nanotubes.\cite{balents97,krotov97,lin98,odintsov99,nersesyan03,kane97} A
recent experiment on carbon-nanotube field-effect devices made from
small-band-gap and nominally metallic carbon nanotubes has shown evidence
for the realization of such a MI state.\cite{deshpande09} Theoretical
work~\cite{balents97,kane97} indicates that carbon nanotubes can be
modeled by the Hubbard model on a two-leg ladder geometry. Since in this
effort we are interested in the generic behavior of a MI in the
non-equilibrium regime, and since  we also need to keep the numerical
effort at a manageable level, we will consider only 1D chains, as opposed
to ladders. Nevertheless our results may set the grounds for future
studies on the appealing two-leg ladder geometry. 

Besides realizations in nanostructures, the electronic properties of some
quasi-one dimensional transition metal oxides are known to be well
described by the one-dimensional Hubbard model.  Most notably, Mott
insulator physics was found to be realized in SrCuO$_2$ and Sr$_2$CuO$_3$
and the specific question of the dielectric breakdown of the MI state was
experimentally addressed by Taguchi {\it et al.}~in Ref.~\onlinecite{taguchi00}.
The actual physics of this experiment, however, may go beyond a simple
Hubbard model description, as has been emphasized by Eckstein, Oka, and
Werner.  \cite{eckstein10a}

An additional and related line of experimental research uses time-resolved 
photoelectron spectroscopy to drive systems with a gap into gapless phases
(see, e.g., Ref.~\onlinecite{perfetti08}). This method allows one
to discriminate Mott insulators from other insulating states.

The outline of the paper is the following. In
section~\ref{modelAndMethods} we present the model and briefly describe
our numerical approach. In section~\ref{results} we present our results
for real-time currents, spin correlations, the double occupancy, and the
entanglement entropy.  Section~\ref{discussion} contains a summary and we
discuss our results, contrasting them against the recent literature.

%%%%%%%%%%%%%%%% Model and Method %%%%%%%%%%%%%%%%%%%%%%%%%%%%%%%%%%%%%%%%%%

\section{\label{modelAndMethods} Model and Methods}

To study the non-equilibrium transport in a Mott insulator we consider a
one-dimensional chain with $L$ electronic sites. The chain is divided into
three different regions: a non-interacting region at the left,
representing a lead; an interacting region in the center, where the Mott
insulator state is located; and another non-interacting region at the
right, representing another lead (see Fig.~\ref{fig:model}). This setup
allows us to include the effects of the leads, complementary to the
approach taken in Ref.~\onlinecite{oka05b}.  The number of sites of the
left (right) lead is $L_{\rm l}$ ($L_{\rm r}$), and in the interacting
region is $L_{\rm int}$. The Hamiltonian of the whole system can be
written as 

\begin{equation}
\label{eq:ham}
H =  H_{\rm int} +  H_{\rm int-leads} + H_{\rm leads},
\end{equation}
where

\begin{eqnarray}
H_{\rm int} &=& -t''\sum^{L_{\rm l}-1+L_{\rm int}} _{\sigma, i = L_{\rm l}+1} 
(c^{\dagger}_{i\sigma}c_{i+1\sigma} + h.c.) 
\nonumber\\& & + \epsilon_0 \sum^{L_{\rm l}+L_{\rm int}}_{\sigma, i =L_{\rm l}+1} n_{i\sigma} 
+ U\sum^{L_{\rm l}+L_{\rm int}}_{i = L_{\rm l}+1} n_{i\uparrow}n_{i\downarrow}
\end{eqnarray}
is the Hamiltonian of a Hubbard chain with onsite Coulomb repulsion $U>0$.
$t''$ is the hopping matrix element between the sites in the interacting region,
and $\epsilon_0$ is the chemical potential in the interacting region.  The
second term in the Hamiltonian is

\begin{eqnarray}
H_{\rm int-leads} &=& -t'\sum_{\sigma} (c^{\dagger}_{L_{\rm l}\sigma}c_{L_{\rm l}+1\sigma} + h.c. \nonumber\\ 
& & +c^{\dagger}_{L_{\rm int}+L_{\rm l}\sigma}c_{L_{\rm int}+L_{\rm
l}+1\sigma} + h.c.),
\end{eqnarray}
connecting the Hubbard chain to the leads with a hopping $t'$, resulting in a
tunneling rate $\Gamma=2{t'}^2$. The third term in the Hamiltonian is

\begin{eqnarray}
H_{\rm leads} &=& -t_{\rm leads}\sum^{L_{\rm l}-1}_{\sigma,i=1} (c^{\dagger}_{i\sigma}c_{i+1\sigma} + h.c.)\nonumber\\ & &
-t_{\rm leads}\sum^{L-1}_{\sigma,i=L_{\rm l}+L_{\rm int}+1}
(c^{\dagger}_{i\sigma}c_{i+1\sigma} + h.c.),
\end{eqnarray}
where $t_{\rm leads}$ is the hopping matrix element in the leads. In most
simulations,  we set $t'=t''$  and we use $t_{\rm leads}=1$ as the unit of
energy unless stated otherwise.  In all the equations above
$c^{\dagger}_{i\sigma}$ represents the creation operator for an electron
at site $i$ and spin projection $\sigma=\uparrow,\downarrow$,
$n_{i\sigma}=c^{\dagger}_{i\sigma}c_{i\sigma}$, and
$n_{i}=n_{i\uparrow}+n_{i\downarrow}$.

We are interested in the time evolution of the MI state in the interacting
portion of the chain when it is driven out of equilibrium by a strong
voltage bias applied between the leads. Therefore, we first need to find
the ground state of the system when the interacting portion of the chain
is at half-filling ($\epsilon_0=-U/2$) and then solve the time-dependent
Schr\"odinger equation for the perturbed system with this state as an
initial condition. The former is accomplished by performing a ground-state
DMRG\cite{white92b, white93, schollwoeck05} calculation with $N=L$
particles.  To perturb the system and to drive the chain out of
equilibrium we add an extra term to the Hamiltonian Eq.~(\ref{eq:ham}),
which has the effect of adding an electric potential at time $t=0$: $$
H_{\mathrm{bias}}= \Theta(t) \sum_{i=1}^L V_i n_i\,, $$ where $\Theta(t)$
is the Heaviside step function and

\begin{equation}
V_i= \left\lbrace 
\begin{array}{ccc}
-V/2 & & i\leq L_l \\
 -(i-L_c)\,E & \mbox{for} & L_l < i \leq L_l+L_{\mathrm{int}}\\ 
 V/2 & & i> L_l+L_{\mathrm{int}}
\end{array}
\right.\,,
\end{equation}
where $L_c=L_l +(L_{\mathrm{int}}+1)/2$. This mimics the effect of an electric field $E=V/(L_{\rm int}+1)$ acting in
the interacting part of the chain, $V$ being the bias voltage induced
between the leads (see Fig.~\ref{fig:model}).

To solve the time-dependent Schr\"odinger equation we use the adaptive
time-dependent DMRG technique~\cite{white04,daley04} with the methods
introduced in Refs.~\onlinecite{alhassanieh06,  dasilva08, boulat08,
kirino08, hm09b,branschaedel10} to simulate non-equilibrium transport.  In
some cases, we use systems with $L_l$ odd and $L_r$ even since we find
that the finite-size effects in the currents are less severe for this
configuration (compare with
Refs.~\onlinecite{alhassanieh06,hm09,branschaedel10} for the case of few
quantum dots).  

The tDMRG simulations are carried out using a third order Trotter-Suzuki
breakup with a time-step of $\delta t=0.1/t_{\mathrm{leads}}$ and under
the constraint of a fixed, maximum discarded weight of $\delta \rho\sim
10^{-7}$. In practice, this implies that one starts the time-evolution with a relatively
small number of states ($m\geq  100$) which then grows fast.  The maximum number of states during the time-evolution  is
$m=1600$ states. Since the accuracy of the numerical results solely
depends on these control parameters, i.e., the discarded weight and the
time step, tDMRG can be considered a quasi-exact method, as the numerical
error can be estimated by varying $\delta t$ and $\delta \rho$. 

In non-equilibrium, the entanglement encoded in the time-dependent wave
function is not bounded by any area law as is the ground-state
entanglement~\cite{eisert10} and may indeed increase extensively as a
function of time. Typically, in so-called global quenches (i.e., the
instantaneous and homogeneous change of one parameter on all sites) one
finds a linear increase of the entanglement entropy (the von-Neumann
entropy) $S_{vN} \sim t$ with time (see, e.g.,
Ref.~\onlinecite{calabrese05} for the case of conformally invariant
systems).  Since the number of states $m$ used in a DMRG calculation
scales as\cite{schollwoeck05} 

\begin{equation} 
m \propto e^{S_{vN}}\,, 
\end{equation}     
reaching long time scales is an exponentially expensive computational task
whenever $S_{vN} \sim t$.  Understanding the time-dependence of $S_{vN}$
itself in generic set-ups is thus an important objective to judge
limitations and capabilities of tDMRG, besides the general and timely
interest in its time-dependence in various kinds of
quenches.\cite{calabrese05,calabrese07,eisler09}

The fact that the number of states increases monotonically with time
defines a maximum time for each simulation as the
time at which the number of states needed to keep the discarded weight
under a fixed value $\delta\rho$ exceeds the maximum of $m=1600$. Then, for representative parameters, 
we
perform several runs with  different $\delta\rho$ to assess and assure the
numerical quality of the data, which ultimately determines the maximum time 
 $t_{\mathrm{max}}$ at which the data for a given observable are still sufficiently reliable.   

We define the symmetrized tunnel current as the average of the two local
currents connecting the interacting region to the left and right leads:

\begin{eqnarray}
\label{eq:j}
j&=& \frac{i t'}{2}\sum_{\sigma} (c_{L_{\rm l},\sigma}^{\dagger}
c_{L_{\rm l}+1,\sigma}-\mbox{h.c.}\nonumber\\ & &+c_{L_{\rm
l}+L_{\rm int},\sigma}^{\dagger}c_{L_{\rm l}+L_{\rm int}+1,\sigma}-\mbox{h.c.}).
\end{eqnarray}
We will  denote the time-dependent expectation value of the symmetrized
current by $J(t)=\langle j(t)\rangle$ whereas the time-averaged current
will simply be denoted as $J$. The currents are measured in units such
that $J/V=2$ corresponds to perfect conductance, i.e., $G_0=2e^2/h$
($e=h=1$ in our work). Local currents $\langle j_i\rangle $ on other bonds
are defined accordingly.

%%%%%%%%%%%%%%%%%%%%%%%%%%%% Results %%%%%%%%%%%%%%%%%%%%%%%%%%%%%%%%%%%%%

\section{\label{results} Results}

The structure of this section is the following. First, we present the
real-time data for the electric current and discuss the properties of the
steady-state currents established after the dielectric breakdown of the
Mott insulator takes place.  Second, we analyze the current-voltage
characteristics.  As the main result of the paper we find a simple
function to describe the current as a function of the bias voltage and the
value of the Lieb-Wu gap associated to the initial Mott insulating state,
similar to the results reported by Oka et al.\cite{oka05b,oka10} Third, we
characterize the current-carrying state in the interacting region by
studying  the time evolution of the charge and the current profiles, the
double occupancy, and the spin-spin correlations. Finally, we discuss the
time-dependence of the entanglement entropy.

\subsection{Real-time data and steady-state currents}
\label{sec:realtime}
\begin{figure}
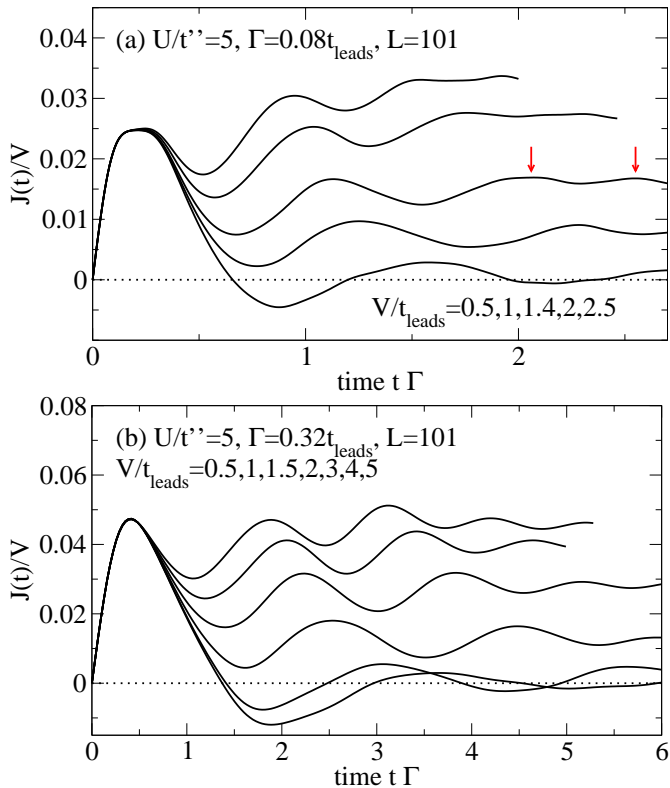

\epsfxsize=0.49\textwidth \centerline{\epsfbox{figure2a.eps}}
\epsfxsize=0.49\textwidth\centerline{\epsfbox{figure2b.eps}}
\caption{\label{fig:real-time-current}
(Color online) Current $J(t)$ as a function of time for $L_{\rm int}=20$, $U/t''=5$,
$t''=t'$,  
 $L=101$ and (a) $\Gamma=0.08t_{\mathrm{leads}}$ (i.e.
$t'=0.2t_{\mathrm{leads}}$),  (b) $\Gamma=0.32t_{\mathrm{leads}}$ (i.e.
$t'=0.4t_{\mathrm{leads}}$).  In (a), $V/t_{\mathrm{leads}}=0.5,1,1.4,2, 2.5$ (bottom to top)
and in (b), $V/t_{\mathrm{leads}}=0.5,1,1.5,2,3,4,5$ (bottom to top). 
Note that the maximum time reached in these
simulations are (a) $t=40/ t_{\mathrm{leads}}$ and (b) $t=20/
t_{\mathrm{leads}}$, in units of the inverse hopping matrix elements in
the leads. The arrows in (a) indicate the time interval used to compute
the steady-state current $J$ for $V=1.4t_{\mathrm{leads}}$.  
} 
\end{figure}

Figure~\ref{fig:real-time-current} shows some examples of the real-time
data for the symmetrized tunnel current obtained from our simulations for
$U/t''=5$ and two values of $\Gamma$.  The transient behavior, in general,
can be expected to depend on both the tunneling rate, set by $\Gamma=2
(t')^2$, and the voltage. For a small interacting region coupled to
non-interacting leads, the transient regime has been studied in
Refs.~\onlinecite{jauho94,schiller00,schmidt08,branschaedel10,hm09b,pletyukhov10}. 
 
In our results, for all voltages, the generic behavior is that the current
first goes through a transient regime, with a maximum reached in the time
window $0\leq t\leq 1/\Gamma $. The figure shows that the time-scale for
reaching the first maximum is  independent of the bias,  while it clearly
depends on $\Gamma$ (this is obvious if one plots the results versus time in units
of $1/t_{\mathrm{leads}}$). Then, accompanied with oscillations whose
period decreases with increasing voltage $V$, we reach a quasi-steady
state regime (typically at times $t\,\Gamma\gtrsim 2...6$) where the
current is constant, apart from oscillations.  The amplitude of the
oscillations decays as the steady-state is approached, yet from our data
we cannot determine whether this decay is an exponential one or not.
The period $t_o$ of the oscillations is a monotonically decreasing function,
similar to the case of single quantum dots in which $t_o\propto 1/V$.\cite{schiller00,pletyukhov10}

The time window over which the steady-state current can be sustained on a
finite system can in principle depend on both $L$ and $L_{\mathrm{int}}$.
$L$ trivially limits the accessible time-scales to $t <
t_{\mathrm{rec}}=2(L-L_{\mathrm{int}})/v_F$, where $v_F$ is the Fermi
velocity in the leads,\cite{alhassanieh06,hm09b,branschaedel10} since by
that time, the perturbations induced in the leads by the application of
the bias have traveled from the interacting region to the boundary and
back, then perturbing the quasi-steady-state currents. $L_{\mathrm{int}}$
does not pose any limit on the stability of the steady-state regime for the setup considered
here, because the bias voltage is introduced \textit{locally} as a
homogeneous electric field. Therefore we choose the values of $L$ and
$L_{\mathrm{int}}$ to give a value of $t_{\mathrm{rec}}$ similar to the
$t_{\mathrm{max}}$ discussed in the previous section,
$t_{\mathrm{rec}}\approx t_{\mathrm{max}}$.

\subsection{I-V characteristics}

In this section we focus on  the steady-state current and its dependence
on the various parameters of the model, presenting results obtained from
extensive numerical calculations.  In practice, we compute the
steady-state current by averaging over one or two periods of the
oscillations at the longest times reached in the simulations (but 
$t<t_{\mathrm{rec}}$) to reduce the effect of the oscillations.  An
example is shown with arrows in Fig.~\ref{fig:real-time-current}(a) for
$V=1.4t_{\mathrm{leads}}$.  We shall find that the current is a simple function of
the bias voltage with all the microscopic details of the model encoded in
the two coefficients  $a$ and $V_c$ in Eq.~(\ref{eq:iv}). I-V curves were
previously presented for the ring geometry for very short chains and in
that case, the currents were extracted from the short-time
dynamics.\cite{oka03}

\begin{figure}
\epsfxsize=0.49\textwidth \centerline{\epsfbox{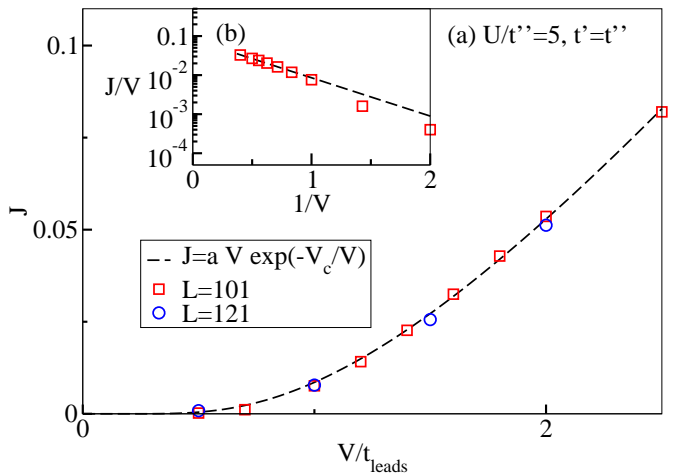}}
\caption{\label{fig:iv}
(Color online) Current-voltage characteristics of the MI. Symbols represent the tDMRG
results for the steady-state current, computed from the data shown in
Fig.~\ref{fig:real-time-current}(a) as explained in
Sec.~\ref{sec:realtime}. Dashed curves are  fits to the function $J(V) =
aVe^{-V_c/{V}}$, $a$ and $V_c$ being free parameters in the fit. (a)
results for $U/t''=5$  with $t'=t''=0.2t_{\mathrm{leads}}$. The plot
includes data from two different system sizes to demonstrate that
finite-size effects are small.  (b) The inset shows the same data on a
log-linear scale for $L=101$.  The agreement with the fit to
Eq.~(\ref{eq:iv}) is excellent (except for very low bias voltages, where
$J$ is of the order of our numerical accuracy).
}
\end{figure}

Figure~\ref{fig:iv} shows the steady-state current $J$
as a function of the bias voltage $V$ for $L_{\rm int}=20$ and 
$U/t''=5$ with $t'=t''=0.2t_{\mathrm{leads}}$. The data from our
numerical simulations for $J$  as a function of the bias voltage fit to
Eq.~(\ref{eq:iv}) with an excellent agreement, $a$ and $V_c$ being the
fitting parameters. Therefore, for values of $V<V_c$ below the  threshold $V_c$, $J$
is exponentially suppressed whereas for values of $V>V_c$ above the threshold,
$J$ increases linearly.  The exponential term is dominant at low bias and
causes the suppression of the current and represents the Landau-Zener
tunneling rate~\cite{zener32} across the Mott gap. The linear term is
dominant at large bias and represents the motion of current-carrying
excitations across the chain in the conducting regime.  

\begin{figure}
\epsfxsize=0.49\textwidth \centerline{\epsfbox{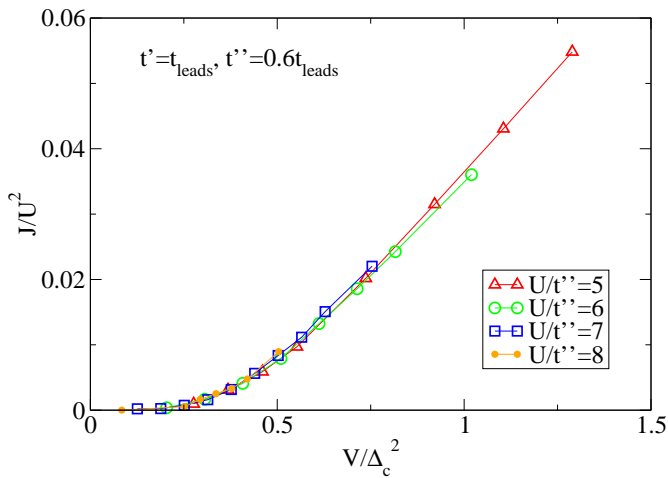}}
\caption{\label{fig:iv3}
(Color online) I-V curves for several $U/t''$ at  fixed values of $t'=t_{\mathrm{leads}}$ 
and $t''=0.6t_{\mathrm{leads}}$. $\Delta_c$ is the charge gap computed
for an isolated  chain of $L_{\mathrm{int}}=20$ from Eq.~(\ref{eq:charge}).  
Symbols are tDMRG data for $L=100$ ($L=80$ for $U/t''=7$). Lines are guides to the eyes.
}
\end{figure}

Figure~\ref{fig:iv3} contains the I-V curves for several different
$U/t''$, keeping $t'$ and $t''$ fixed. Motivated by Fig.~2 from
Ref.~\onlinecite{oka03}, we have plotted the steady-state current as a
function of $V/\Delta_c^2$, where $\Delta_c$ is the charge gap.  We have
calculated the charge gap for finite systems  with $L_{\mathrm{int}}=20$
sites, not connected to any leads, using
\begin{equation}
\Delta_c= \lbrack E_0(N+2,S^z)+E_0(N-2,S^z)-2E_0(N,S^z) \rbrack /2\,,
\label{eq:charge} \end{equation}
where $E_0(N,S^z)$ is the ground state energy in subspaces with $N$
fermions and a total spin projection $S^z$.  Using this, and by also
plotting the current in units of $U^2$, all curves collapse on a single
one, which, in particular, suggests $V_c \propto \Delta_c^2 $, as expected
for a Landau-Zener type of breakdown of the MI state.\cite{oka03,oka05b}
As we show here, this important fingerprint of Landau-Zener physics also
survives upon coupling the interacting region to leads.

We here therefore find  essentially the same dependence of $V_c$ on $U$ as Oka 
{et al.},\cite{oka03,oka05b} namely $V_c\propto \Delta_c^2$, but
with incorporating the leads into the model. There are some differences, though. First, it
should be noted that our time-averaged current is extracted from
simulations that reach much longer times than Ref.~\onlinecite{oka03}
where only the short-time dynamics was available to estimate the
steady-state currents.  Second, we do not find an abrupt increase of the
current at the threshold voltage, in contrast to Ref.~\onlinecite{oka03}. Therefore, our data are in a better
agreement with the result of mapping the problem to a quantum walk (see
Fig.~3 in Ref.~\onlinecite{oka05c}).  We attribute the quantitative
differences between Fig.~5 in Ref.~\onlinecite{oka03} and our
Fig.~\ref{fig:iv} to the difference in the calculation of $J$, the fact
that our systems are larger, and the inclusion of the leads. 

\begin{figure}
\epsfxsize=0.49\textwidth \centerline{\epsfbox{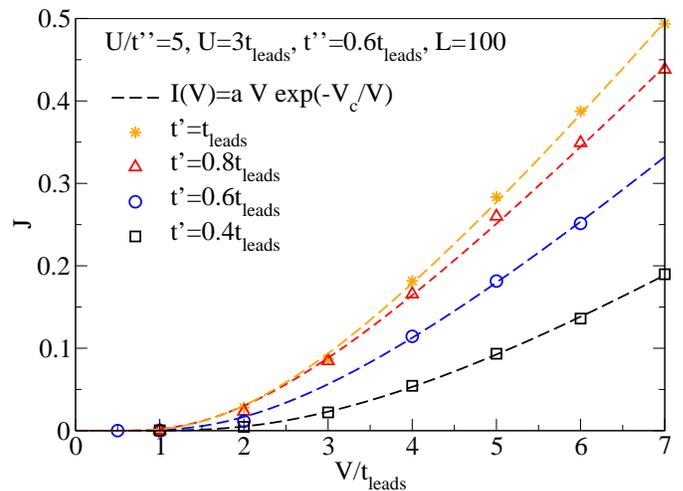}}
\caption{\label{fig:iv2}
(Color online) I-V curves for $t'\not= t''$ at a fixed $U/t''=5$ with
$t''=0.6t_{\mathrm{leads}}$.  $t'/t_{\mathrm{leads}}=0.4,0.6,0.8,1$
(bottom to top). Dashed curves are  fits to the function $J(V) =
aVe^{-V_c/{V}}$, with $a$ and $V_c$ being free parameters in the fit. 
}
\end{figure}

\begin{figure}[b]
\epsfxsize=0.49\textwidth \centerline{\epsfbox{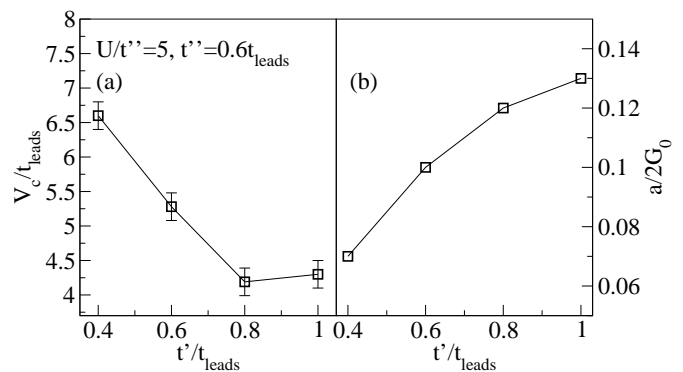}}
\caption{\label{fig:factors}
(Color online) (a) Threshold voltage $V_c$ vs. $t'$; (b) prefactor $a$ in Eq.~(\ref{eq:iv})
vs. t'. $U/t''=5$, $t''=0.6t_{\mathrm{leads}}$, $L=100$.
}
\end{figure}

To further explore the effect of the leads on $V_c$ we have computed I-V
curves for a fixed value of $U/t''=5$ and several $t'$, as shown in
Fig.~\ref{fig:iv2}.  Qualitatively, a larger $t'$ leads to an overall
increase of the current as reflected in the $t'$-dependence of $a$ to be
discussed later on. The threshold exhibits a weak dependence on $t'$ as well,
as we demonstrate in Fig.~\ref{fig:factors}(a). Our observation is that
$V_c(t'<t'')>V_c(t'=t'')$ and $V_c(t'>t'')<V_c(t'=t'')$. The latter behavior 
can be explained by the observation that close to the interface, the local
charge gap depends on $t'$: $t'<t''$ leads to a slightly enhanced gap
compared to the bulk gap and vice versa. As a consequence, the double
occupancy $\langle d_i\rangle=\langle n_{i\uparrow}n_{i\downarrow}\rangle
$ (discussed in detail below) in the interacting region is enhanced close
to the interface compared to the bulk value for $t'>t''$, while it is
suppressed for $t'<t''$.  Therefore, for $t'<t''$ the contacts suppress
the current, giving rise to an increase of $V_c$. In the case of 
 $t'>t''$, the largest local gap is in the bulk of the MI and decreases towards
the boundary. The decrease of $V_c$ as $t'\to t_{\mathrm{leads}}$ can 
be understood as a consequence of a smaller mismatch between $t',t''$ 
and $t_{\mathrm{leads}}$ in that limit, which should give rise to a increase in the
transmission of electrons across the interface region.
Note that we observe that boundary effects in
the initial state typically decay to the bulk values over a distance of
about 5 sites, suggesting that $L_{\mathrm{int}}=20$ is a reasonable
choice to probe both the bulk and contact properties.

Next, we address the dependence of the prefactor $a$ on $t'$.  The
coefficient $a$ sets the value of the differential conductance in the
conducting regime. We present our results for $a$ and various combinations
of $t'$ in Fig.~\ref{fig:factors}(b), in units of $G_0=2e^2/h$.
Interestingly, in all cases studied, $a<2G_0$. Moreover, this coefficient
$a$ monotonically increases with $t'$ or $\Gamma=2t'^2$.  To summarize,
$a$ depends on both $t'$ and $U$ and, phenomenologically, we find that
$a\propto U^2$ results in a convincing collapse of the I-V curves for
$U> 4t''$ (compare Fig.~\ref{fig:iv3}).

We have also studied the dependence of the I-V curves on
$L_{\mathrm{int}}$ (not shown in the figures). We find that 

\begin{equation}
V_c\propto (L_{\rm int}+1)\Delta_{c}^{2} \quad \mbox{and} \quad a \sim 1/(L_{\mathrm{int}}+1)\,.
\label{eq:b}
\end{equation}
This suggests that the breakdown should be viewed as {\it field}-driven
with $E=V/(L_{\mathrm{int}}+1)$ taking the role of the electric field. We
may therefore rewrite Eq.~(\ref{eq:iv}) as: \begin{equation} 
J=\tilde a E \,
\mbox{exp}(-E_c/E)\,. \label{eq:field} 
\end{equation}
 This interpretation is in agreement with
Refs.~\onlinecite{oka03,oka05b,eckstein10a}, and we stress that the functional form of the I-V curve 
described by Eq.~(\ref{eq:field}) holds despite the presence of the leads. As we have shown here, the effect
of the leads is a small deviation of  $\tilde a$ and the threshold field $E_c$ from the bulk values
(compare Fig.~\ref{fig:factors} and Refs.~\onlinecite{oka03,oka05b}).

%%%%%%%%%%%%%%%%%%%%%%%%%%%%%%%%%%%%%%%%%%%%%%%%%%%%%%%%%%%%%%%%%%%%%%%%%%%%%%%%%%%%%%%%%%%%%%%%
\subsection{Characterization of the current-carrying state}

The goal of this section is to characterize the current-carrying state in
the interacting region. To  this end, we measure the electronic density
and electronic current density profiles in the interacting region, the
average double occupancy, and also the spin-spin correlations, yielding 
 the spin structure factor.
%%%%%%%%%%%%%%%%%%%%%%%%%%%%%%%%%%%%%%%%%%%%%%%%%%%%%%%%%%%%%%%%%%%%%%%%%%%%%%%%%%%%%%%%%%%%%%%%

\subsubsection{Density and current profiles}

\begin{figure}
\epsfxsize=0.49\textwidth \centerline{\epsfbox{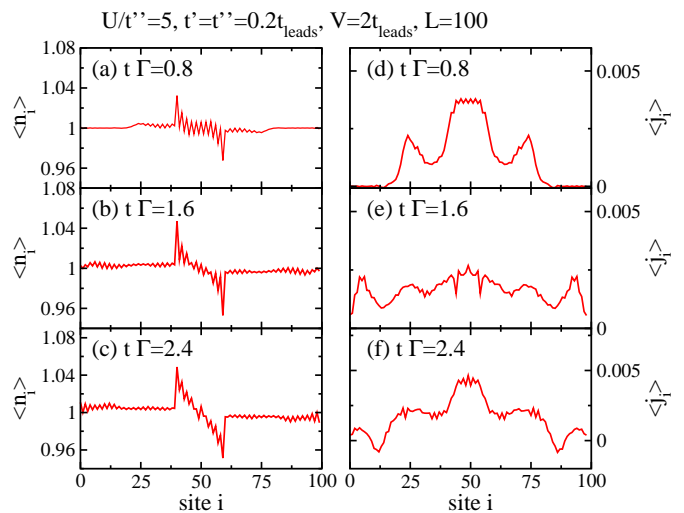}}
\caption{\label{N-x.L.151.U.1.0.V.2.0}
(Color online) (a)-(c) Charge density $\langle n_{i}\rangle$ as a function of position at
different times $t\, \Gamma =0.8,1.6,2.4$ for $U/t''=5$,
$t'=t''=0.2t_{\mathrm{leads}}$, $V = 2t_{\mathrm{leads}}$, and $L=100$.
(d)-(e) Profile of the local currents $\langle j_i\rangle $ at the same
times as in panels (a)-(c).
} 
\end{figure}

Figure~\ref{N-x.L.151.U.1.0.V.2.0}(a)-(c) show the charge density $\langle
n_i\rangle$ as a function of position at different times for $U/t''=5$,
$V=2t_{\mathrm{leads}}$, and $L_{\rm int}=20$ and the corresponding local
currents $\langle j_i\rangle$ in (d)-(f).  In the steady state, the charge
in the interacting portion of the chain has a linear profile following the
profile of the applied bias.  The overall charge density in the
Hubbard chain remains at half-filling. 

From the results for the local currents, we see that the currents take
finite values on all sites, which actually happens immediately after
applying the potential. This clearly distinguishes the breakdown mechanism
induced by a linear profile from other spatial forms of the bias voltage.
For instance, in the simplest case in which $V_i=0$ in the interacting
region and $V_i=\pm V/2$ in the left(right) lead, the physics underlying
the breakdown is quite different as we have verified in additional
simulations (results not shown here).  In this case, the redistribution of
the charge inside the interacting region can be described as an effective
doping of the MI region from the two interfaces. This implies that the
bulk of the interacting region will experience the effects caused by
turning on bias with a delay, set by the length of the interacting region.

Turning back to Fig.~\ref{N-x.L.151.U.1.0.V.2.0}, to justify that the
steady state in an extended system has been reached, the currents need to
be constant both in time and space. From Fig.~\ref{N-x.L.151.U.1.0.V.2.0},
we see that $\langle j_i \rangle =\mbox{const}$ is not fulfilled, although
the charge flow in and out of the system is constant, apart from the
relatively small oscillations discussed before (compare
Fig.~\ref{fig:real-time-current}). This suggests that for the time scales
reached in our simulations, the interacting region still undergoes a
reorganization of charges and local energies. Indeed, from the data of
Fig.~\ref{N-x.L.151.U.1.0.V.2.0}, we find $\langle j_i-j_{i-1}\rangle
\not= 0$, even at $t\,\Gamma \sim 2.5$.

%%%%%%%%%%%%%%%%%%%%%%%%%%%%%%%%%%%%%%%%%%%%%%%%%%%%%%%%%%%%%%%%%%%%%%%%%%%%%%%%%%%%%%%%%%%%%%%%

\subsubsection{Double occupancy}

\begin{figure}
\epsfxsize=0.49\textwidth \centerline{\epsfbox{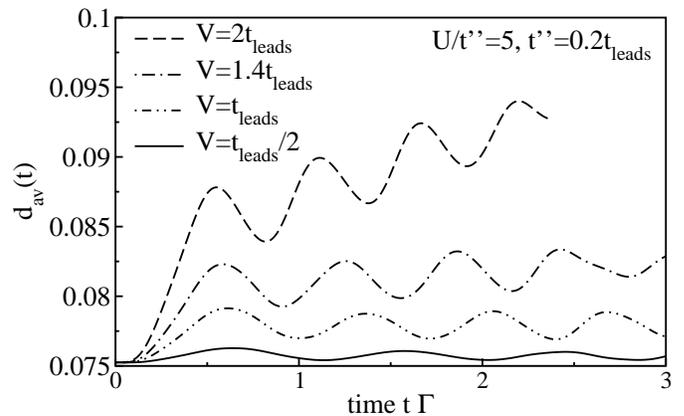}}
\caption{\label{fig:double-occupancies}  
(Color online) Average double occupancy $d_{\mathrm{av}}(t)$ in the interacting region as
a function of time for $U/t''=5$, $V/t_{\mathrm{leads}}=0.5,1,1.4,2$, and
$t'=t''=0.2t_{\mathrm{leads}}$.  }
\end{figure}

Figure~\ref{fig:double-occupancies} shows the average double occupancy per
site in the interacting portion of the chain \begin{equation}
d_{\mathrm{av}}(t) = \frac{1}{L_{\mathrm{int}}}
\sum_{i=L_l+1}^{L_L+L_{\mathrm{int}}} \langle d_i(t) \rangle \end{equation} as a
function of time for $U/t''=5$ and $V/t_{\mathrm{leads}}=0.5,1,1.4,2$.  At
all these voltages, the average double occupancy oscillates with a period
given by $t_o=t_o(V)$ that decreases with increasing voltage $1/V$, similar to the behavior of the currents.

Depending on the bias voltage two different behaviors can be observed. For
bias voltages below the threshold $V<V_{c}$, i.e., in the regime of
exponentially suppressed currents, $d_{\mathrm{av}}(t)$ is essentially
constant, apart from the oscillations. For bias voltages above the
threshold $V>V_c$, $d_{\mathrm{av}}(t)$ increases according to
\begin{equation}
d_{\mathrm{av}}(t)= A+B\,t+C\,\cos(D\,t)\label{eq:dav}\,,
\end{equation}
i.e., linearly in time after averaging over the period $t_{o}=2\pi/D$.
The slope $B$ can be interpreted as the rate of the production of pairs of
doublons and vacancies induced by the effective electric
field.\cite{oka05b,oka10}  Quite notably, the double occupancy never
saturates over the time window simulated, i.e., a steady-state regime for
this quantity is not reached in our simulations, even if the system is in
the steady-state regime for the tunneling current.  A similar observation
has been made in the DMFT study by Eckstein \textit{et
al.},\cite{eckstein10a} who also report a monotonically increasing double
occupancy $d_{\mathrm{av}}(t)$ in the steady-current regime.
They ascribe this to the 
fact that the work done by the field is proportional to  $j\,E$, which in
a regime of constant currents is a constant. Hence this increase in energy
has to go into the internal energy of the MI, in the absence of any 
dissipation or leads.

We shall here elaborate in more detail on this reasoning, adopting it to
our set-up that includes the leads. To explain the time-dependence of $d_{\mathrm{av}}$ of Eq.~(\ref{eq:dav}) 
we exploit the fact that the equation of motion for the \textit{average} double
occupancy operator $\hat{d}_{\mathrm{av}}=(1/L_{\mathrm{int}})
\sum_{i=L_l+1}^{L_L+L_{\mathrm{int}}}n_{i\uparrow}n_{i\downarrow}$ is the same as
the one for the interaction energy. After some straightforward algebra,
one gets
\begin{equation} \label{eq:double_occups_motion}
\frac{d}{dt} \hat{d}_{\mathrm{av}}=\frac{1}{UL_{\mathrm{int}}}\left(
-\frac{d\hat{T}}{dt}+E\sum_{i=L_l+1}^{L_l+L_{\mathrm{int}}-1}j_{i}\right),
\end{equation}
where $\hat{T}=\hat{T}_{\mathrm{int}}+\hat{T}_{\mathrm{int-leads}}$ is the
kinetic energy operator involving sites at the interacting region, and $E$
is the constant electric field. For times in the steady-current regime,
the time integration of the second term in the RHS gives a linear
dependence on time, as the current is approximately constant. Assuming
that $d_{\mathrm{av}}$ is small, as Fig.~\ref{N-x.L.151.U.1.0.V.2.0}
suggests, we can expand the quantum mechanical average of the kinetic
energy operator in the interacting region as $\langle
\hat{T}_{\mathrm{int}}\rangle\approx
T_{0}+\epsilon_{\mathrm{d}}d_{\mathrm{av}}+\mathcal{O}(d_{\mathrm{av}}^{2})$,
where $T_{0}$ is the kinetic energy of the filled lower Hubbard band, and
$\epsilon_{\mathrm{d}}$ is the kinetic energy of a doublon. As a filled
band cannot increase its kinetic energy, the time derivative approximates
as $d\langle \hat{T}_{\mathrm{int}}\rangle/dt\approx
\epsilon_{\mathrm{d}}d d_{\mathrm{av}}/dt$.  With this assumption one can
move the contribution from $\hat T_{\mathrm{int}}$ to the LHS of
Eq.~(\ref{eq:double_occups_motion}) and conclude that the time derivative
of the average double occupancy is
\begin{equation}
\label{eq:double_occups_motion_rev}
\frac{d}{dt}{d}_{\mathrm{av}}(t)\propto
-\frac{d}{dt}{T}_{\mathrm{int-leads}}+E \sum_{i=L_l+1}^{L_l+L_{\mathrm{int}}-1} j_{i}+\mathcal{O}(d_{\mathrm{av}}(t)^{2})
\end{equation}
where all operators have been substituted by their quantum mechanical
averages, and we have  changed the equality in
Eq.~(\ref{eq:double_occups_motion_rev}) to a proportionality to
accommodate the term stemming from the kinetic energy of the doublons. The
first term in the RHS is the energy flowing out of the interacting region
carried away by the particles transfered to the leads. If the interacting
part is an isolated system as in Ref.~\onlinecite{eckstein10a}, this term
is absent.  The interpretation of Eq.~(\ref{eq:double_occups_motion_rev}) is
that although the establishment of the steady-current regime implies a
linear increase of the double occupancy and therefore of the interaction
energy, part of this energy is transfered to the leads when accelerated
particles leave the interacting region. This reduces the rate at which the
double occupancy increases, allowing the system to stay in the
steady-current regime for a longer time.  The increase in the double
occupancy implies that the system is not in a true steady-state in the
sense that there are observables that depend on time.

As for the existence and the nature of a true steady-state, two scenarios are conceivable.
Obviously, due to the bounded spectrum, the increase of the double occupancy
cannot go on forever, so eventually it has to saturate. An extreme case
would be that $d_{\mathrm{av}}$ takes its maximum value $d_{\mathrm{max}}=0.5$ compatible with
the system being at half filling on average. Consequently, the
current would vanish in this case. Alternatively, the internal energy
could saturate at some time, reflected in $d_{\mathrm{av}}=\mbox{const}<0.5$
(where the RHS is the maximum possible value assuming an average half filling
of the interacting region). In that case, a finite current flow would be possible
and the energy gain due to particles getting accelerated by the electric field
would have to be balanced by an equal energy flow into the leads.
In either case, the reorganization of doublons may take longer than the time
needed to reach the steady-state regime for the current. In particular, it is
well known that the dynamics of doublons in one-dimensional systems with $U>W$
where $W$ is the bandwidth can be slow, if not even  delayed by metastable regimes (see, e.g., 
Refs.~\onlinecite{alhassanieh08,dasilva10,hm09a} for 1D systems and Refs.~\onlinecite{rosch08} for higher dimensions). This aspect has also been touched
upon in Ref.~\onlinecite{eckstein10a}. 

Unfortunately, our simulations
are restricted in the accessible  times, and we can thus not clarify this
point, leaving it as an open question for future research.

%%%%%%%%%%%%%%%%%%%%%%%%%%%%%%%%%%%%%%%%%%%%%%%%%%%%%%%%%%%%%%%%%%%%%%%%%%%%%%%%%%%%%%%%%%%%%%%%

\subsubsection{Spin-spin correlations}

\begin{figure}
\epsfxsize=0.49\textwidth \centerline{\epsfbox{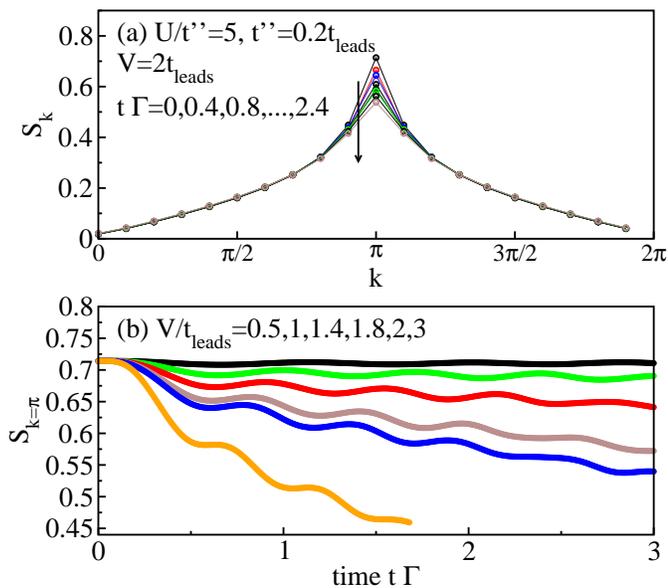}}
\caption{\label{fig:spin}
(Color online) (a) Spin structure-factor in the interacting region as a function of time
for $U/t''=5$ ($t'=t''=0.2t_{\mathrm{leads}}$) and $V=2t_{\mathrm{leads}}$.
(b) $S_{k=\pi}$ in the interacting region as a function of time for
$V/t_{\mathrm{leads}}=0.5,1,1.4,1.8,2,3$ (top to bottom).
}
\end{figure}

The (longitudinal) spin structure factor can be  computed from the
spin-spin correlations by taking a Fourier transform ($i,j\in \lbrack
L_{l}+1, L_{l}+L_{\mathrm{int}}\rbrack$):

\begin{equation}
S_k = \frac{1}{L_{\mathrm{int}}} \sum_{l,m} e^{-i(l-m)k} \langle S_l^z S_m^z\rangle\,.  
\end{equation}
Figure~\ref{fig:spin}(a) shows the spin structure-factor at different
times for $U/t''=5.0$ and $V=2t_{\mathrm{leads}}$.  The main feature is
the survival of antiferromagnetic correlations in the current-carrying
state: the shape of the spin structure-factor remains qualitatively the
same, yet the weight of the $k=\pi$  instability decreases steadily with
time. We therefore show $S_{k=\pi}(t)$ for several bias values $V$ in
Fig.~\ref{fig:spin}(b) as a function of time.  The figure unveils that,
similar to the case of the average double occupancy, a steady-state regime
for this observable is not reached in our simulations, i.e., on the
longest times reached and for the system sizes considered here.  Similar
to the linear  {\it increase} of the average double occupancy, the
$S_{k=\pi}(t)$ {\it decreases} linearly in time.

%%%%%%%%%%%%%%%%%%%%%%%%%%%%%%%%%%%%%%%%%%%%%%%%%%%%%%%%%%%%%%%%%%%%%%%%%%%%%%%%%%%%%%%%%%%%%%%%

\subsection{Entanglement entropy}

The entanglement entropy is defined as

\begin{equation}
S_{vN,x} = - \mbox{tr}\lbrack\rho_x \mbox{ln}(\rho_x) \rbrack,
\end{equation}
where $\rho_x$ is the reduced density matrix of a block of the length $x$
(counting from the left end of the chain). The reduced density matrix and
its spectrum of eigenvalues is a key object in DMRG and the
entanglement entropy is thus one of the easiest accessible
quantities.\cite{schollwoeck05}  

Let us begin by recalling some established analytical results on the
entanglement growth in quantum quenches in systems with conformal
invariance: In a {\it global} quench (i.e., the change of a parameter on
all sites), $S_{vN,x}\propto t$ (Ref.~\onlinecite{calabrese05}) whereas in
a {\it local} quench, $S_{vN,x} \propto
\mbox{ln}(t/t_0)$.\cite{calabrese07,eisler07} For the case of a global
quench, this has been confirmed in numerous numerical calculations, mostly
using DMRG (see, e.g., Refs.~\onlinecite{laeuchli08,dechiara06}).  

\begin{figure}
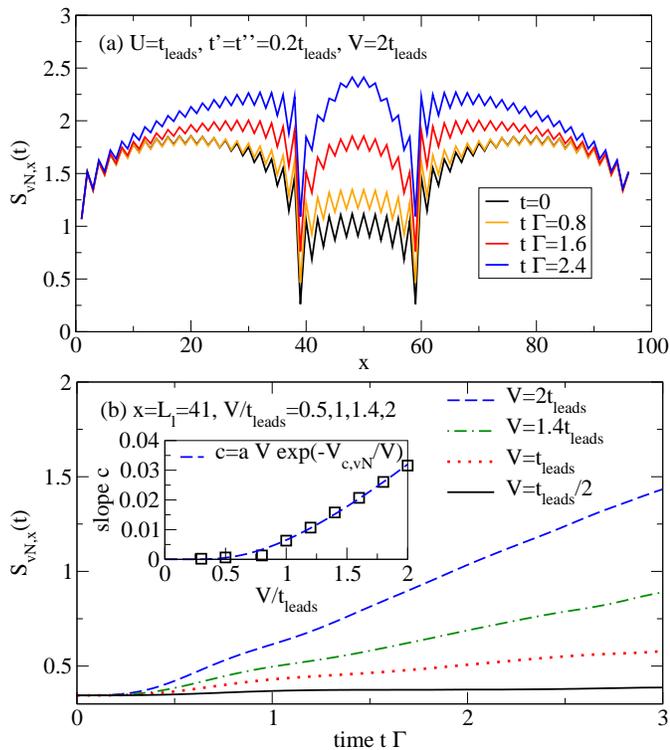

\epsfxsize=0.49\textwidth \centerline{\epsfbox{figure10a.eps}}
\epsfxsize=0.49\textwidth \centerline{\epsfbox{figure10b.eps}}
\caption{\label{fig:svn}
(Color online) (a) Entanglement entropy $S_{vN,x}(t)$ vs. position $x$ of the cut taken
in the bipartition for times $t\,\Gamma=0,0.8,1.6,2.4$.  (b) $S_{vN,x}(t)$
vs. time $t$ for $x=L_l=41$. This cuts the system across the left link that
connects the left lead to the interaction region
($V/t_{\mathrm{leads}}=0.5,1,1.4,2$, from bottom to top ). Inset: slope
$c$ of $S_{vN,x}(t)=c\,t $ computed in the time-interval $t\,\Gamma\in
\lbrack 0.25,3\rbrack$. The dashed line is a fit to
$c=a\,V\,\mbox{exp}(-V_{\mathrm{c,vN}}/V)$.  
In both panels, $U/t''=5$, $t'=t''=0.2t_{\mathrm{leads}}$.}
\end{figure}

Our situation is different, since a parameter - the bias voltage - is
changed on all sites, but with an explicit site dependence. Our results
for $S_{vN,x}=S_{vN,x}(t)$ are displayed in Fig.~\ref{fig:svn}. Panel (a)
shows $S_{vN,x}=S_{vN,x}(t)$ vs. $x$ for all possible cuts accessed in a
DMRG run for a fixed value of $V=2t_{\mathrm{leads}}$ at different times.
The overall increase of $S_{vN,x}$ as a function of time is evident.

The key question here is how the flow of particles in the conducting
regime gives rise to an increased entanglement between, say, the left lead
and the rest of the system. In particular, we expect basically no increase
in the insulating regime of bias voltages $V<V_c\propto \Delta_c^2$. To
address this point, we plot $S_{vN,x}(t)$ with $x=L_l$ in
Fig.~\ref{fig:svn}(b) for several bias voltages. Generally, we find that
$S_{vN,x}=c\, t $. The dependence of the prefactor $c$ on bias voltage $V$
is shown in the inset of Fig.~\ref{fig:svn}(b): its dependence on $V$ can
be described by the same functional form as the tunnel current, namely: 
\begin{equation}
c\propto V \,\mbox{exp}(-V_{c,vN}/V)\,. 
\end{equation} 
In particular, we find that
$V_{c,vN}\approx V_c$ within the accuracy of our numerical simulations,
where $V_c$ is the threshold voltage extracted from Fig.~\ref{fig:iv}.
This is consistent with the picture that entanglement is predominantly
induced by propagating particles, in contrast to global quenches, in which
$\langle n_i(t)\rangle=$const.

While the observation of $S_{vN,x}\propto t$ implies that the simulations
carried out here become exponentially costly at long times, we note that
in similar set-ups, namely the case in which a confining potential of a
linear form is present in the {\it initial} state and its removal at $t=0$
is used to drive the time-evolution, a weaker logarithmic increase is
found. Specifically, for the exactly solvable $XX$ model, Eisler et al. 
 report $S_{vN,x}\propto \mbox{ln}(t)$.\cite{eisler09} Two main
differences between their set-up and ours need to be pointed out. First,
in our case, the application  of the bias $V_i$ destroys the MI state {\it
and} drives the current flow.  Conversely, in the set-up of
Ref.~\onlinecite{eisler09}, the initial state already has an inhomogeneous
particle density, implying that correlations in the initial state are
already very different from the respective ground state ones at the same
filling. These open questions and observations call for a full analysis of
the behavior of $S_{vN,x}$ in global quenches with site dependent changes
of parameter, that we leave as a future project.

\section{\label{discussion} Discussion}

In this paper we have studied the dielectric breakdown of a Mott insulator
state in a realistic model with an interacting chain connected to
non-interacting leads. Our numerical results confirm that the steady-state
current as a function of the applied voltage is, over a wide range of
voltages, described by a simple universal function, with  all the
microscopic details of the model encoded in two coefficients related to
the conductance in the metallic regime and the value of the threshold
voltage. Our work further elucidates the influence of contacts to the
leads on the I-V curve: the overall current is a monotonically increasing
function in the inverse tunneling rate $1/\Gamma$ and the threshold, on
finite systems, also exhibits a weak dependence on the contacts. 
%%%%%%%%%%%%%%%%%%%%%%%%%%%%%%%%%%%%%%%%%%%%%%%%%%%%%%%%%%%%%%%%%%%%%%%%%%%

The dielectric breakdown of the one-dimensional Hubbard model was studied
under dissipative tunnelling into the environment introduced by a
imaginary gauge potential in Ref.~[\onlinecite{fukui98}], and upon the
application of a strong electric field introduced by a gauge potential in a
ring geometry in Refs.~[\onlinecite{oka03, oka05a, oka05b, oka10}]. The
main conclusion of the latter papers is that the dielectric breakdown of
the Mott insulator can be understood in the same terms as the one in band
insulators, with the only change that  the band gap has to be substituted for with the Mott gap in
the calculation of the Landau-Zener parameter (i.e., the threshold field).
The time-averaged current in small Hubbard rings shows a collapse of the
currents to a universal curve when the currents are plotted as a function
of the Landau-Zener parameter,\cite{oka03} sharing the same qualitative
traits as our Fig.~\ref{fig:iv}, with a negligible current before the
breakdown and a linear I-V characteristics at biases larger than the
threshold. An important conclusion of our work is the confirmation that
the mechanism of the dielectric breakdown corresponds to the Landau-Zener
tunneling mechanism and this mechanism survives upon coupling the
interacting region to leads.

It should be noted that another very recent tDMRG study by Kirino and
Ueda\cite{kirino10} has adressed the destruction of the MI state upon
application of a strong voltage as well. There are important differences with our
work, though. In Ref.~\onlinecite{kirino10}, no leads are included, and the
bias is applied as a step-function function to a homogeneous MI, measuring
the local current on the central link. While the I-V curve also shows an
activated behavior, it is not clear whether the MI is also destroyed
through a Landau-Zener mechanism in the set-up of Kirino and Ueda. In
particular, they report $V_c \propto \Delta_c$, in contrast to the results
by Oka et al.~and ours (compare Fig.~\ref{fig:iv3}). This illustrates the
rich and various physical scenarios that can underlie the breakdown of an
insulating state, depending on the way the bias is applied.
%%%%%%%%%%%%%%%%%%%%%%%%%%%%%%%%%%%%%%%%%%%%%%%%%%%%%%%%%%%%%%%%%%%%%%%%%%%
   
We have also studied the conducting state that is reached after the
breakdown. The spin-spin correlations remain antiferromagnetic in the
steady state. A decrease in the amplitude of the correlations is observed
as the bias exceeds the threshold value. The conducting state can also be
distinguished from the initial insulator by an increase in the double
occupation. In other words, the electric field creates excitations as
pairs of doublons and holons that can carry the current.\cite{kluger91}
The production rate of these excitations should be reflected in the
production rate of doubly occupied sites. Quite notably, the
time-dependence of both the double occupancy and the spin-spin
correlations implies that the interacting region is not in a true steady
state yet, in which these quantities would become stationary as well.

Finally, we have also computed the time-dependence of the entanglement
entropy. This quantity increases linearly with time in the conducting
regime, implying that tDMRG simulations become exponentially expensive at
long times. On the positive side, studying transport through single
quantum dots or extended structures has qualitatively the same
computational complexity, since in both cases, $S_{vN,x}\propto t$
(unpublished results for one quantum dot, see Ref.~\onlinecite{hm09b}).
Therefore, going from single to many quantum dots is equally feasible with
this method, in contrast to other state-of-the-art techniques such as
time-dependent NRG \cite{anders08} or real-time QMC.\cite{werner10} In the
former, the complexity scales with the dimension of the interacting region
and in the latter approach, the dynamical sign-problem is expected to
become more severe for structures more complex than a  single quantum dot.
We have here demonstrated that tDMRG can successfully be applied to
compute I-V curves of extended systems, complementing our earlier work on
non-equilibrium transport in the single-impurity
problem.\cite{alhassanieh06,dasilva08,hm09b}

While our numerical analysis of several properties of the current-carrying
state should be helpful in better understanding its properties, we
acknowledge that a more intuitive picture of the non-equilibrium
steady-state is still desirable. For instance, one would like to contrast
the current-carrying steady-state against effective ground-state reference
systems, an approach which in certain non-equilibrium cases works quite
well.\cite{hm08a} Moreover, the interesting concept of an effective
temperature, often used in studies of quantum quenches with a relaxation
into a thermalized state (see Ref.~\onlinecite{cazalilla10} and references
therein), should be further explored for current-carrying states.

In conclusion, we have shown that the dielectric breakdown of the Mott
insulator can be understood in terms of the Landau-Zener mechanism using a
realistic setup that matches the experiment since we include the leads. Furthermore we have been able
to fully characterize the steady-state currents as a function of the bias
voltage with a simple form, covering the whole range of voltages and
microscopic parameters, that can be experimentally tested. 

\begin{acknowledgments}

We thank  M. Daghofer, H. Onishi, G. Roux and D. Schuricht  for very useful discussions. I.G.
acknowledges support from MICINN through grant FIS2009-13520. A.E.F.
thanks NSF for support through grant DMR-0955707. E.D. is supported by the
Division of Materials Science and
Engineering, Office of Basic Energy Sciences, U.S. Department of Energy.

\end{acknowledgments}

%%%%%%%%%%%%%%%%%%%%%%%%%%%%%%%%%%%%%%%%%%%%%%%%%%%%%%%%%%%%%%%%%%%%%%%%

\end{document}